# SBML Qualitative Models: a model representation format and infrastructure to foster interactions between qualitative modelling formalisms and tools

Claudine Chaouiya*[1], Duncan Berenguier#[2], Sarah M Keating#[3], Aurelien Naldi#[4], Martijn P. van Iersel[3], Nicolas Rodriguez[3,5], Andreas Dräger[6,7], Finja Büchel[7], Thomas Cokelaer[3], Bryan Kowal[8], Benjamin Wicks[8], Emanuel Gonçalves[3], Julien Dorier[9], Michel Page[10,11], Pedro T. Monteiro[1,12], Axel von Kamp[13], Ioannis Xenarios[9], Hidde de Jong[10], Michael Hucka[14], Steffen Klamt[13], Denis Thieffry[15], Nicolas Le Novère[3,5], Julio Saez-Rodriguez*[3], Tomáš Helikar*[8]

*Corresponding Authors
# Equal contributors

Affiliations:

1. Instituto Gulbenkian de Ciência, Rua da Quinta Grande 6, 2780-156 Oeiras, Portugal
2. Institut de Mathématiques de Luminy, Campus de Luminy, Case 907,13288 Marseille Cedex 9, France
3. European Bioinformatics Institute (EMBL-EBI), European Molecular Biology Laboratory, Wellcome Trust Genome Campus, Hinxton, Cambridge CB10 1SD, United Kingdom
4. Center for Integrative Genomics, University of Lausanne, CH-1015 Lausanne, Switzerland
5. The Babraham Institute, Babraham Research Campus, Cambridge, CB22 3AT, United Kingdom
6. Department of Bioengineering, University of California San Diego, La Jolla, California, 92093-0412, United States of America
7. Center for Bioinformatics Tuebingen (ZBIT), University of Tuebingen, 72076 Tübingen, Germany
8. Department of Mathematics, University of Nebraska at Omaha, Omaha, NE, United States of America
9. Swiss-Prot & Vital-IT group, SIB- Swiss Institute of Bioinformatics, Center for Integrative Genomics, University of Lausanne, Quartier Sorge - Batiment Genopode, CH-1015 Lausanne, Switzerland
10. INRIA Grenoble – Rhône-Alpes, 655 avenue de l'Europe, Montbonnot, 38334 Saint-Ismier Cedex, France
11. IAE Grenoble, Université Pierre-Mendès-France, Domaine universitaire BP 47, 38040 Grenoble Cedex 9, France
12. Instituto de Engenharia de Sistemas e Computadores - Investigação e Desenvolvimento (INESC-ID), Rua Alves Redol 9, 1000-029 Lisbon, Portugal
13. Max Planck Institute for Dynamics of Complex Technical Systems, Sandtorstr. 1, D-39106 Magdeburg, Germany
14. Computing and Mathematical sciences, California Institute of Technology, Pasadena, CA 91125, United States of America
15. Institut de Biologie de l'Ecole Normale Supérieure (IBENS) - UMR CNRS 8197 - INSERM 1024 46 rue d'Ulm, 75230 Paris Cedex 05, France

## Abstract

**Background:** Qualitative frameworks, especially those based on the logical discrete formalism, are increasingly used to model regulatory and signalling networks. A major advantage of these frameworks is that they do not require precise quantitative data, and that they are well-suited for studies of large networks. While numerous groups have developed specific computational tools that provide original methods to analyse qualitative models, a standard format to exchange qualitative models has been missing.

**Results:** We present the System Biology Markup Language (SBML) Qualitative Models Package ("qual"), an extension of the SBML Level 3 standard designed for computer representation of qualitative models of biological networks. We demonstrate the interoperability of models via SBML qual through the analysis of a specific signalling network by three independent software tools. Furthermore, the cooperative development of the SBML qual format paved the way for the development of LogicalModel, an open-source model library, which will facilitate the adoption of the format as well as the collaborative development of algorithms to analyze qualitative models.

**Conclusion:** SBML qual allows the exchange of qualitative models among a number of complementary software tools. SBML qual has the potential to promote collaborative work on the development of novel computational approaches, as well as on the specification and the analysis of comprehensive qualitative models of regulatory and signalling networks.

## Background

Studies by S. Kauffman [1] and R. Thomas [2] founded the logical discrete approach to model biological molecular networks and analyse their behaviours. In these networks, components (e.g., genes or proteins) assume discrete values representing their activity levels (e.g., gene expression). Components are connected by directed edges that embody regulatory (causal) effects, forming an influence network. The activity level of each component evolves depending on the activity levels of the components influencing it. The rules that determine component activity levels are defined in terms of logical rules or functions, corresponding to the underlying biological/biochemical regulatory mechanisms. The dynamical behaviour of the network is then generated by evolving the component levels following a specific updating scheme (e.g., synchronous, asynchronous or stochastic). The dynamics can subsequently be represented in terms of a state transition graph, where the nodes represent (discrete) states of the model, while the edges denote transitions between these states.

While other mathematical frameworks, including differential equations, can be used to model biological processes in great detail, the logical formalism is particularly suitable for the modelling of large networks for which precise kinetic data are not available. In fact, logical models have become increasingly popular. They have been recently used to model complex dynamical behaviours and provide insights into numerous biological systems, including **gene regulatory networks** (e.g., [3–6]), **signal transduction** (e.g., [7–14]), as well as **cell cycle** (e.g., [15–18]), in species ranging from **bacteria** and **viruses** (e.g., [3, 19, 20]) to **yeast** (e.g., [17, 21–23]), **flies** (e.g., [24–26]), **plants** (e.g., [27, 28]), and even to **humans** (e.g., [11, 12, 18, 29]).

Often based on qualitative knowledge of regulatory mechanisms and published data, discrete models can be assembled through a "bottom-up" approach, whereby each logical function represents specific, biological interactions between the components of the network. Recently, "top-down" approaches have also emerged as a means of constructing logical models by automatic inference from high-throughput experiments (e.g., [30]).

Many simulation and analysis software tools for logical models already exist, including ADAM [31], BoolNet [32], BooleanNet [33], Cell Collective [34, 35] CellNetAnalyzer [36], CellNOpt [30] , ChemChains [37], GINsim [38], Odefy [39], SimBoolNet [40], SQUAD [41], etc.

The state transition graphs describing the discrete dynamics of networks may be huge and therefore difficult to analyze. This has led several groups to propose the use of model-checking techniques [42] to explore properties of these graphs, in terms of attractors and paths leading to those attractors [43]. A number of logical modelling tools allow properties of the state transition graphs to be verified by means of existing model-checking tools, such as NuSMV [44–47]. The properties are formulated in terms of temporal logic or in a suitable high-level query template capturing recurrent biological questions [48]. The model checker tests if the state transition graph, which may be explicitly generated or implicitly encoded in a symbolic description of the model, satisfies the property. For example, while GINsim exports symbolically encoded logical models to SMV files, BIOCHAM integrates NuSMV [47] providing an interface for the specification and verification of properties expressed in several temporal logics [46]. A detailed description of the use of model-checking techniques in the context of qualitative models of biological networks is outside the scope of this paper, but see reference [49] for a review and additional examples.

Over the years, different formats have been developed to store logical models, ranging from simple text files containing truth tables and/or logical functions to XML-based file formats. Standards such as the Systems Biology Markup Language (SBML, [50]) or the Systems Biology Graphical Notation (SBGN, [51]) have been developed to enable unified exchange of biological/biochemical molecular maps. SBML supports process-based mathematical frameworks with a reaction-centred description of biochemical processes. Because the building blocks of qualitative models are fundamentally different from species and reactions used in (core) SBML models, previous attempts to represent logical models in SBML led to a distorted use of the standard. Indeed, variables in Boolean networks, logical models and some Petri nets represent discrete levels of activities rather than amounts/numbers of molecules. Consequently, the processes involving them cannot be described as reactions per se, but rather as transitions between states.

The specification of SBML (Level 3) is modular and thereby enables the development and inclusion of packages providing additional features. Using this modular structure, we developed a novel Qualitative Models ("qual") package to support the standard definition and exchange of qualitative (discrete) models.

It is worth noting that, although SBML qual development mainly focused on logical models, standard Petri Nets can also be encoded in this new format due to commonalities between the frameworks. Indeed, while Petri nets are mostly used to study metabolic networks, they have also been employed to model regulatory and signalling networks (see reviews [52–54]). Currently, the Petri net community relies on specialized exchange formats (e.g., PNML, http://www.pnml.org) and simulation tools that support SBML core (e.g., [55]).

Given the open source nature of SBML qual and the collaborative nature of the SBML community, the new standard should be swiftly adopted and implemented in most existing tools supporting logical models and their relatives such as Petri nets and hybrid models. The cooperation on SBML qual further fostered synergistic efforts to articulate and improve existing tools, leading to the launching of the Common Logical Modelling Tools (CoLoMoTo) project (http://co.mbine.org/colomoto/), which gathers many groups developing and using logical modelling software tools.

# Methods

## Development of the qual package
A draft proposal of a SBML package to encode qualitative models was initially proposed in 2008. Between 2008 and 2012, the proposal was refined, through community consultations and dedicated meetings by developers of various related software tools, and in particular members of the CoLoMoTo project. In 2011, the proposal was accepted through a community vote. The final specification was accepted by the SBML Editors in the spring of 2013.

## LibSBML & JSBML
LibSBML is an application programming interface (API) library for reading, writing, manipulating and validating content expressed in the SBML format [56]. It is written in ISO C and C++, provides language bindings for .NET, Java, Python, Perl, Ruby, MATLAB and Octave, and includes many features that facilitate the adoption and use of both SBML and the libSBML. JSBML, a pure Java library for SBML, provides an API that maps all SBML elements to a flexible and extended Java type hierarchy whilst striving for 100% compatibility with the libSBML Java API [57]. Both libraries provide support for SBML qual in their development branches (as of August 2013), and will include support in their next major releases. LibSBML and JSBML are freely available as source code and binaries for all major operating systems under the LGPL open source terms (see http://sbml.org/Downloads). JSBML has been integrated in the LogicalModel library (see Results).

## Computer simulations

To demonstrate the interoperability of models via SBML qual, we analyzed a specific signalling network using three different software tools briefly described below.

**CellNOpt** is an open-source software used for creating logic-based models of signal transduction networks [30]. CellNOpt consists of a set of R packages available in Bioconductor, which are also available via a Python wrapper, as well as a Cytoscape plug-in (CytoCopteR) which contains a SBML qual importer and exporter. CellNOpt converts a network (a signed, directed graph) into a scaffold of all possible models compatible with the network and subsequently trains this scaffold with data [58]. It includes a variety of formalisms: (i) Boolean models, simulated via synchronous update or by computation of steady-states, (ii) semi-quantitative constrained Fuzzy logic, and (iii) ordinary differential equations (ODEs) derived from the logical model [30]. While the choice of a specific formalism depends on the data at hand, scope, and question, the followed workflow is similar. The network can be simplified by compressing nodes that are intermediates between perturbed or measured nodes. Links impinging on nodes that are not observable (with no readout downstream) or not controllable (with no perturbation upstream of them) are also taken aside as their status cannot be derived from the data.

CellNOpt generates logical models as hyper-graphs by adding all combinations of OR and AND gates that are compatible with the network (i.e., Sums of Products; [59]). This leads to a hyper-graph representing a superposition of all Boolean models compatible with the initial network. Subsequently, an optimisation procedure is applied to find the combination of gates and the parameters that best explain the data, by minimizing an objective function that quantifies the difference between data and simulation, while penalizing model size. This provides an optimum model or, more generally, a family of optimal models. Optimization can be performed using a built-in genetic algorithm, or using external optimization packages; in particular CellNOpt is connected to Meigo [60]. Furthermore, CellNOpt can leverage Answer Set Programming to efficiently find all possible Boolean models via the software package caspo [61].

Once an optimal model (or family of models) has been generated, it can be analyzed in various ways. For example, it can be simulated to predict the outcome of new experiments [58]. One can also analyze the properties of a family of models, or compare models obtained for different cell types [62]. One can also identify missing links in the network using the module CNORFeeder [63]. The flexibility of the scripting languages (R, or Python) simplifies the writing of analysis workflows, and Cytocopter enables combined analysis with other Cytoscape tools and plug-ins.

**GINsim** is a free (Java) software application devoted to the logical (multi-valued) modelling of regulatory and signalling networks [38, 64]. It provides a user-friendly graphical interface to define models from scratch. Models can also be imported from different formats. GINsim supports the simulation of logical models and generates the resulting state transition graphs, considering a range of update policies (see below). GINsim also offers a number of functionalities to explore the dynamical properties of logical models, some of which (e.g., determination of stable states) can be efficiently analyzed without generating the complete network dynamics.

It is well known that regulatory circuits (or feedback loops) can generate crucial dynamical properties [65]: positive circuits (encompassing an even number of inhibitions) produce multi-stability whereas negative circuits (encompassing an odd number of inhibitions) underlie stable oscillations. To help analyze these properties, GINsim identifies all the regulatory circuits embedded into a network and compute the regions of the state space, called functionality contexts, where they generate the related property (multi-stability *versus* oscillations).

One can use various updating schemes to generate the dynamics of a logical model. When in a given state, several components are called to change their values, these updates can be done synchronously, asynchronously, or considering a priority scheme [15]. Under the synchronous scheme, all components are updated simultaneously, leading to one transition at most for each state and thus resulting in a deterministic (linear) state transition sequences. Under the asynchronous scheme, single component updates are considered separately, assuming that underlying delays are different but unknown; consequently, alternative trajectories are often generated, giving rise to a nondeterministic state transition graph.

Of particular interest is the asymptotical dynamical behaviour of these models, which is captured by the notion of *attractors.* From a logical point of view, attractors take two forms: stable states, and terminal cyclic strongly-connected components (as defined in graph theory). Note that stable states and terminal elementary cycles (where in each state, a unique component is updated) are shared between synchronous and asynchronous updating schemes but this is not the case for other cyclic attractors. GINsim supports both the synchronous and asynchronous updating schemes, which can lead to rather distinct dynamical properties. In particular, asynchronous dynamics can be quite complex.

In this respect, Hierarchical Transition Graphs (HTG) provide a compact and informative view of the dynamics in the form of a graph where nodes embody sets of states that are either irreversible (denoting irreversible sequences of states) or strongly connected (denoting oscillations in the form of transient or terminal complex components). For more details on HTG, see reference [66].

Finally, to handle the analysis of large models, several groups have devised reduction methods [10, 58, 67, 68]. In this respect, the last (beta) version of GINsim allows users to get rid of (pseudo-) output species that do not regulate other nodes or regulate only pseudo-output nodes. This reduction has no impact on the number, nature and reachability of the attractors and it is particularly efficient for signalling networks as shown with our example model (see Results section).

**The Cell Collective** is a web-based platform for the construction, simulation, and analysis of Boolean-based models [34, 35]. The platform includes a Knowledge Base for users to annotate the models and keep track of experimental research papers associated with each interaction included in the model. Within the platform, models can be shared directly on the web or via download using the SBML qual format (as well as in the form of text files including the list of logical expressions, and as .csv files with truth tables).

Models constructed in Cell Collective are Boolean (each species has a Boolean function associated with it, and assumes either an active or inactive state), and simulations can also include stochastic elements. Furthermore, data input/output from the analyses are continuous, providing a semi-quantitative measure to better match modelling results with laboratory experiments [14, 37]. At the input level, this is accomplished by assigning a probability of being active in time $t$ to each external species (i.e., those with no regulators), in contrast to classical Boolean simulations where each external species is fixed to 0 or 1.

The activity level, or the probable active state of an output species, is measured by calculating the ratio of 0's and 1's over the last $n$ time steps ($n$ is configurable to any discrete value, [35, 37]); this ratio (multiplied by 100) provides the activity level on the $y$-axis (e.g., Figure 5; these parameters can be changed through the user interface). In the case of real-time simulations, as a single simulation evolves in time, the activity level of each species in the model is calculated as the ratio of 0's and 1's within a predefined sliding window [35, 37].

One of the assets of the Cell Collective is its user interface, which has been carefully designed to enable the construction of computational models in a nontechnical fashion, in order to render modelling also amenable to nonmodellers. That is, the construction of the models is based purely on provided qualitative knowledge about a particular regulatory mechanism (e.g., kinase X phosphorylates and activates species Y), without the need to manually enter Boolean expressions (these are created in the background based on the biological data provided [69]). Although creating relatively small Boolean models can be easily done by writing Boolean functions, defining models with species regulated by many regulators, or through complex regulatory mechanisms can often result in complex, nested functions (e.g., [14, 29]), which can be cumbersome to define manually even for seasoned modellers.

# Results

In this section, we present the SBML qual package and its validation by exchanging and interpreting a moderately complex signalling network model among three independent software tools. In addition, we illustrate the interest of model exchange by applying complementary simulation and analysis features. The section ends with a description of the LogicalModel library.

## The SBML qual package

The SBML qual package extends the core SBML Level 3 standard, and enables standard exchange and interoperability of discrete (logical) models. The full specification is available at http://identifiers.org/combine.specifications/sbml.level-3.version-1.qual.version-1.release-1 [70]. The structure of SBML qual is depicted in Figure 1. The rationale of the format relies on the features of qualitative models, with a (discrete) state space and event-driven state transition processes.

The main elements of an SBML qual document are *QualitativeSpecies,* representing the entities of the model as the molecular components of the network, and *Transitions*, which contain the logical rules defining the state of given species at each iteration step.

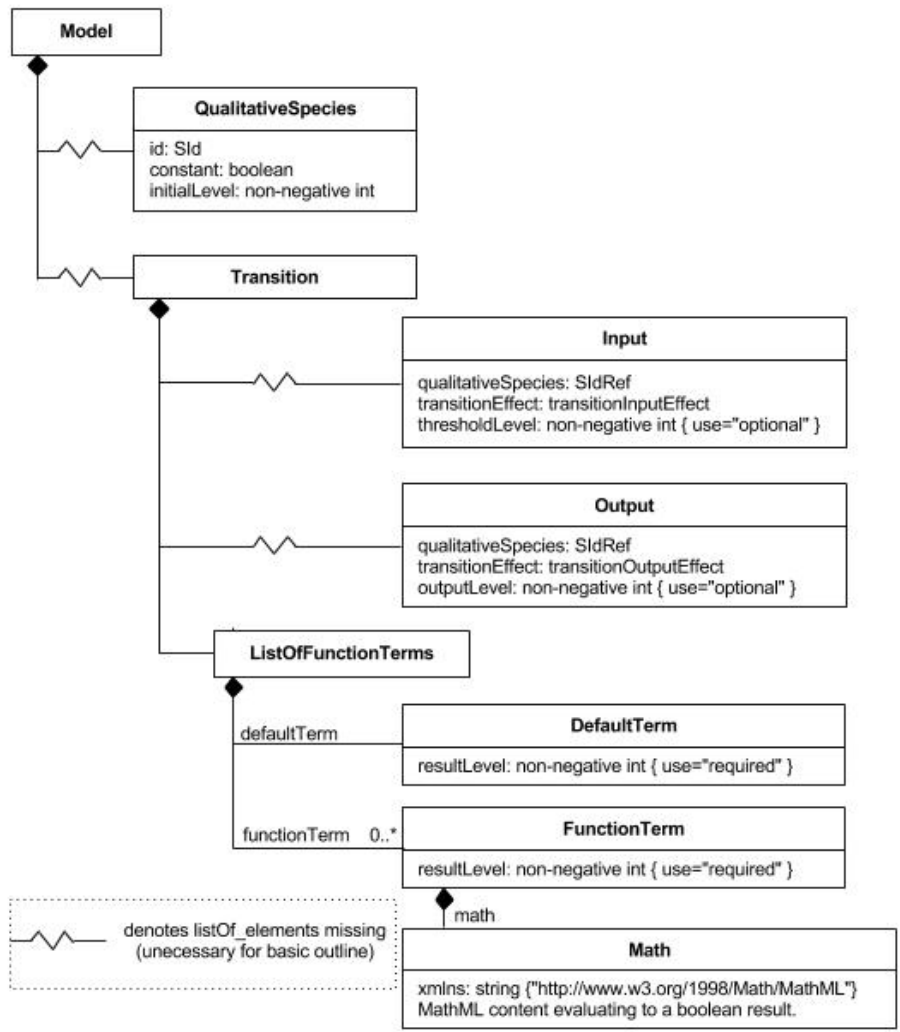

**Figure 1. Reduced UML diagram that captures the SBML L3 Qualitative Models package (SBML qual).** The *QualitativeSpecies* represent the entities involved in the model. These are referenced as either *Inputs* or *Outputs* of the *Transition* element. A *Transition* describes the way the level of each *QualitativeSpecies* may be altered depending on the levels of other entities in the model (see the full UML diagram in [70]).

Each *QualitativeSpecies* assumes a discrete value (e.g., 0 or 1 for the Boolean case), and its definition bears an attribute *initialLevel* that specifies the value(s) at the beginning of the simulation, and an attribute *maxLevel* that specifies the maximal level allowed. For instance, *maxLevel* would be 1 in a Boolean model. As for the *Species* of SBML Core, a *QualitativeSpecies* is associated with a compartment.

A *Transition* comprises *Inputs* (*QualitativeSpecies* mentioned in the logical function), *Outputs* (species whose values at time *t*+1 are determined by the logical rules evaluated at *t*) and *FunctionTerms,* containing conditions, as well as the values that the *Output* species will assume at time *t*+1, whenever a given condition is met. At each time step, all *FunctionTerms* within a *Transition* are evaluated. The term evaluating to *true* dictates the resulting state and the *Output* species are updated accordingly at time *t*+1.

Each member of the *ListOfFunctionTerms* associated with a *Transition* contains a mathematical expression that returns a Boolean, as well as a *resultLevel* that indicates the level to be applied to the *Outputs* when this expression evaluates to true. A *defaultTerm* is also defined to establish the result when none of the *FunctionTerms* apply. The combined set of *defaultTerm* together with the list of *FunctionTerms* establish the state transitions for the entities involved. Figure 2 provides an illustration of a simple Boolean model encoded in SBML qual.

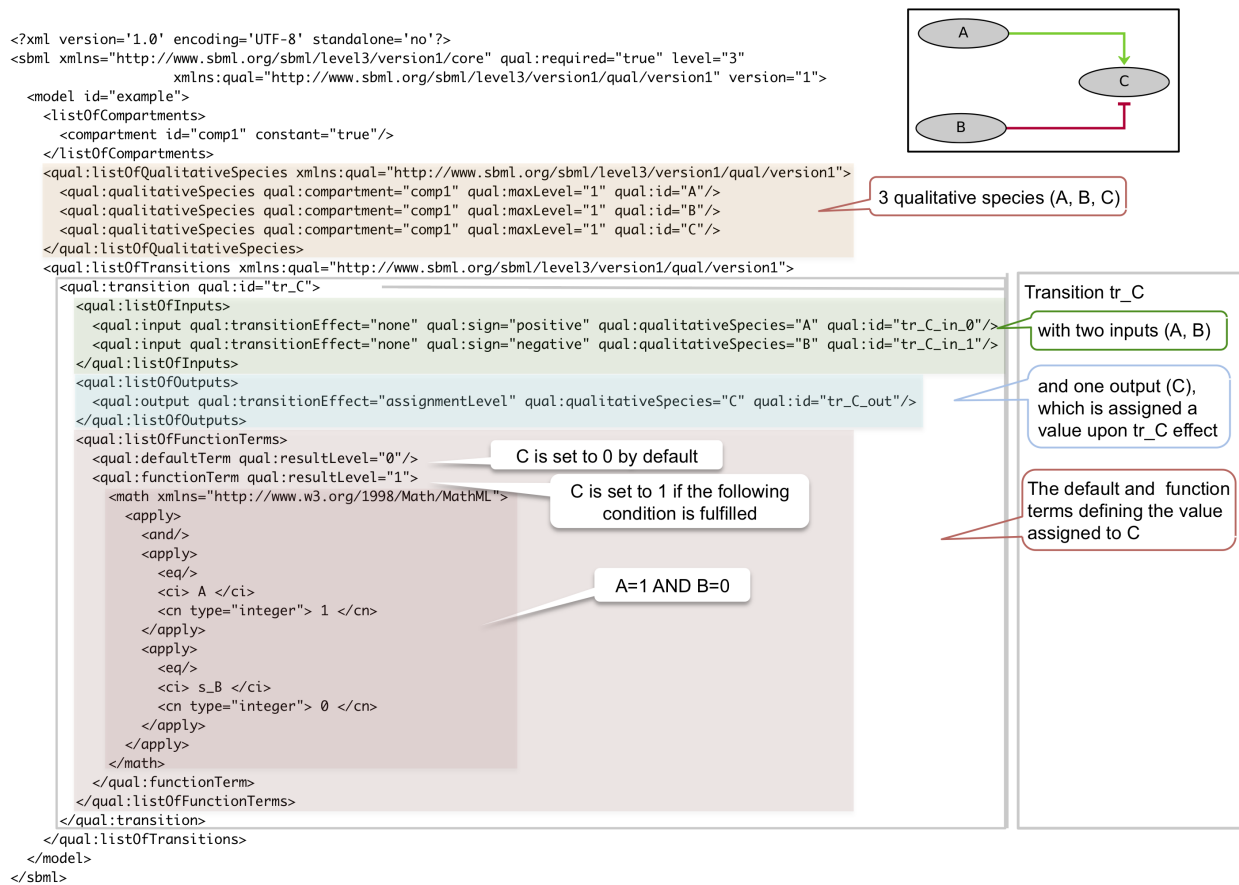

**Figure 2:** A simple Boolean network encoded in SBML Qual; the three components are Boolean and C is activated by A and inhibited by B. These regulatory effects are embodied in the *Transition* element (tr_C), which has 2 *inputs* (A and B, which levels are not modified by the transition), one *output (C,* whose assigned level is defined in the *listOfFunctionTerms* of the transition). The *defaultTerm* is set to 0, while its sole *functionTerm* specifies (in the form of a MathML element) that C is 1 when A=1 and B=0. Note that the SBML format is intended to be readable by computers only; hence the code presented in this figure is for illustration purposes only.

## Demonstration of model interoperability

As part of the SBML extension development and approval process (by the SBML Editors) is the requirement that at least two independent software tools fully implement the proposed package. CellNOpt, GINsim, and the Cell Collective have been recently registered with the SBML community as tools currently supporting the *SBML qual* package. These tools have been further used to demonstrate how a logical model can be handled with different software tools using SBML qual as an exchange format.

As the three aforementioned software tools can provide different perspectives on the dynamics of discrete/logical models, this section is organized to demonstrate their complementarity. More specifically, we present a conceptual pipeline that enables scientists to derive a discrete model from high-throughput data, conduct thorough analyses, and ultimately use the model to further guide experiments. CellNOpt is used to derive a logical model via a top-down approach, exploiting experimental high-throughput data and an initial qualitative description of a signalling network (see Methods). Inferred models are subsequently simulated and analysed using additional techniques implemented in GINsim and the Cell Collective.

### Generation of a EGF/TNFα discrete model with CellNOpt

The focus of CellNOpt is to utilize experimental data to generate logical models based on prior knowledge on signalling pathways (Prior Knowledge Networks, PKNs). The example model used herein is based on a PKN that combines two important mammalian signalling pathways, induced by the Epidermal Growth Factor (EGF) and Tumour Necrosis Factor alpha (TNFα). EGF and TNFα ligands stimulate ERK, JNK and p38 MAPK cascades, the PI3K/AKT pathways, and the NFκB cascade. In addition, the network encompasses cross-talks between these pathways, as well as two negative feedback loops: one in the NFκB cascade and one in the MAPK cascade. Note that this network was previously used in [71] to illustrate a variety of logic modelling approaches using synthetic data. Here, however, we slightly modified this PKN by adding an autocatalytic feedback loop in the phosphatase (ph) regulating the activation of SOS-1 (Figure 3).

The PKN was subsequently trained using the synchronous update Boolean simulation (CNORdt package), in combination with the CNORFeeder package to obtain the optimal logical model (see Methods) used as an example in this paper (Figure 3). Instead of using experimental data, an ordinary differentiation equation (ODE) model representing the "true network" was employed to generate the data and train the PKN. These data (in the form of time series) were thus obtained by simulating the ODE model upon stimulation of EGF and TNFα, and inhibition of PI3K and Raf-1 in different combinations. The readout nodes (i.e., the proteins whose activities were measured upon stimulation) are highlighted in Figure 3. To reflect imprecisions in our knowledge of biological pathways, the topology of the data generator model ("golden standard") is slightly different from the PKN. More precisely, a link from Map3K7 to MKK7 has been omitted in the PKN, to which an extra edge from PI3K to Map3K1 was further added. A workflow with CellNOpt was able to recover the "golden standard" model from this PKN and the experimental data. This final model was then exported to SBML qual and simulated and analyzed using all three tools.

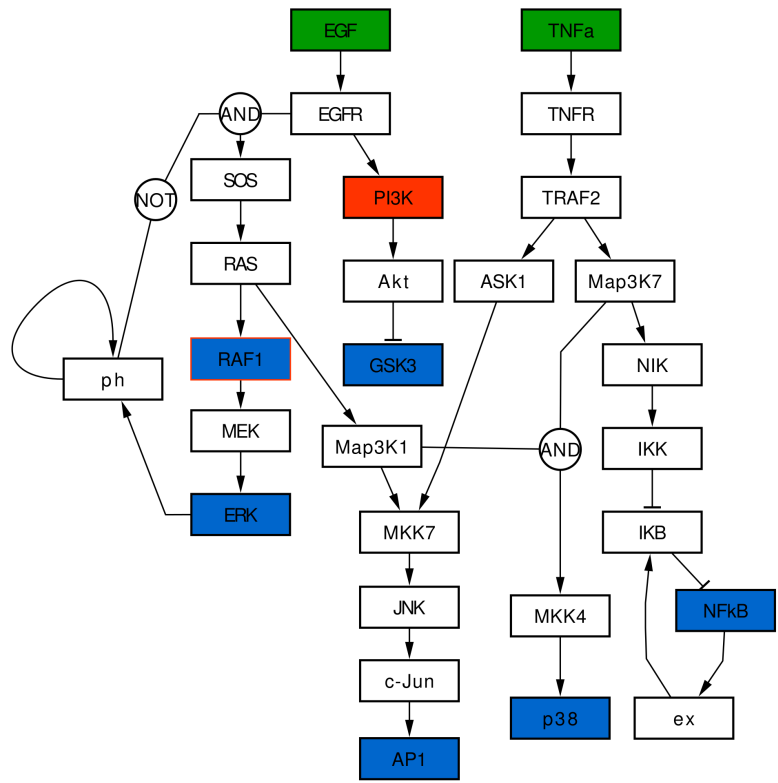

**Figure 3: Boolean model obtained by CellNOpt and vizualized using the Activity Flow language of the Systems Biology Graphical Notation [51] and drawn with CySBGN [72].** Different colours define the experimental design of the data used to train the model: (i) green boxes denote external stimuli, (ii) red boxes correspond to species blocked by kinase inhibitors, and (iii) blue boxes denote species that were measured (readouts).

### Dynamical cross-validation of the model

Dynamical (synchronous) simulations of the EGF/TNFα network for the four different initial conditions gave consistent results with the three software tools. In this respect, Figure 4 shows the consistent global state evolution, attractor reachability, as well as temporal evolution of selected nodes for two of these conditions (see Figure S1 for a full set of simulation results). Depending on the initial condition, simulations result in one of two stable states or in one of two cycles encompassing six states.

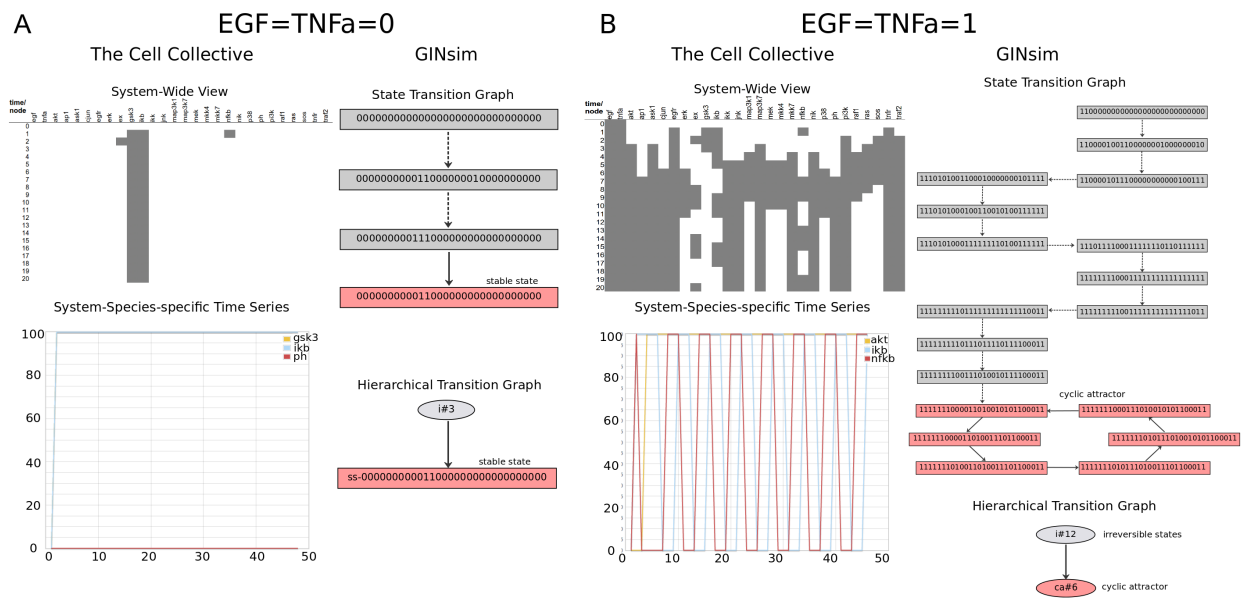

**Figure 4. Dynamic profile of the EGF/TNFα model.** The model was simulated with consistent results in the Cell Collective (left column of panels A & B), GINsim (right column of panels A & B), and CellNOpt (data not shown, but simulations were consistent with those presented here). Synchronous simulations were performed under four input conditions: (i) EGF=TNFα=0; (ii) EGF=TNFα=1; (iii) EGF=0 & TNFα=1; (iv) EGF=1 & TNFα=0. Results for conditions in (i) and (ii) are presented in this figure (the remaining two can be found in Figure S1). Charts at the top of the Cell Collective column correspond to the overall dynamic profile across all nodes in the model. Black cells correspond to active (1) states, whereas inactive (0) states are white. The bottom graphs in the Cell Collective column illustrate the time course of selected nodes. The GINsim columns show State Transition Graph and the Hierarchical Transition Graph (HTG) generated with the tool. Note that due to the synchronous updating, the irreversible components of the (HTG) correspond to linear chains of states. A) EGF=TNFα=0. The network reaches a steady state (shown in both GINsim and the Cell Collective column) after 3 transient states. The order of the individual species states in the steady state generated by GINsim is sorted in the same (alphabetical) order, as presented in the Cell Collective column. B) EGF=TNFα=1. After 12 transient states, the network reaches a cyclical attractor encompassing six states. Note that in order to simulate the example model in the Cell Collective as a traditional Boolean network (i.e., with binary input/output), the external species were set to 100 or 0, and the sliding window was set to 1 (see Methods).

## Model analyses in GINsim

As mentioned in the previous section, starting from the null state, simulations using synchronous updating of the example model result in a unique attractor for each of the four combinations of the two external inputs (EGF and TNFα), either a stable state or a simple terminal (Figure S1). Although we expect to get the same attractors under the asynchronous update, their reachability may be affected. Moreover, the number of states possibly visited before reaching an attractor may greatly differ between synchronous and asynchronous simulations, as well as the characteristics of the transient dynamics.

For example, when EGF=1 and TNFα=1, the asynchronous state transition graph is substantially larger with about 116k states, as opposed to the 19 states obtained with a synchronous update (Figure 4B). To contain the size of the state transition graph, we can reduce the model by removing all (pseudo-) output nodes (cf. Methods). Applying this reduction and thereby eliminating 11 nodes as (pseudo-) outputs results in a significant reduction of the state transition graph (down to 546 states, Figure 5A). Importantly, the unique reachable attractor is identical to that obtained with the synchronous update. The resulting HTG (Figure 5A) has a peculiar staged structure with a series of irreversible components (labeled i# followed by the number of states included) that the system may leave to undergo transient oscillations (labeled #ct), to eventually reach the final cyclic attractor (labelled #ca). Interestingly, by defining priority classes and imposing that IKK update is slower than all other species, one can get rid of all these transient oscillations (Figure 5B).

To maintain input components (EGF and TNFα) constant, implicit self-activations are defined. These two (functional) positive circuits explain the presence of at least four attractors; the combinations of input values define a partition of the state space in four disconnected regions. Using GINsim, we can verify that the EGF/TNFα model encompasses two additional functional circuits: a three-element negative circuit involving IκB, NFκB and ex, and a positive auto-regulatory circuit on ph. The functionality context of the negative circuit corresponds to IKK=1 (which is the case when TNFα is 1). This negative circuit enables the attractors where IκB, NFκB and ex oscillate. The functionality context of the positive auto-regulatory circuit is defined by ERK=0. This circuit explains the presence of two attractors when EFG=0.

For the input configuration where EGF=0 and TNFα=1, starting from an initial state with ERK=1, under the asynchronous update, the system is able to reach two cyclic attractors that differ by the presence of ph (Figure 5C). Trajectories leading to the terminal cycle with ph=0 are discarded by the synchronous update in which the decrease of ERK (in the absence of EGF and thus MEK1) occurs together with the increase of ph, already in the first step of the simulation, leading to the cyclic attractor where ph=1.

Using GINsim, common perturbations such as gene knock-outs or ectopic gene expressions, as well as their effects, can be easily simulated. For instance, knocking out IKK eliminates oscillations (that were present under wild-type simulations when TNFα= 1), as a result of the interruption of the IκB-NFκB-ex negative circuit (Figure 5D). Similarly, the simulation of ERK ectopic expression, disrupting the functionality context of ph auto-regulation, leads to the loss of the multi-stability when EGF=0 (Figure 5C).

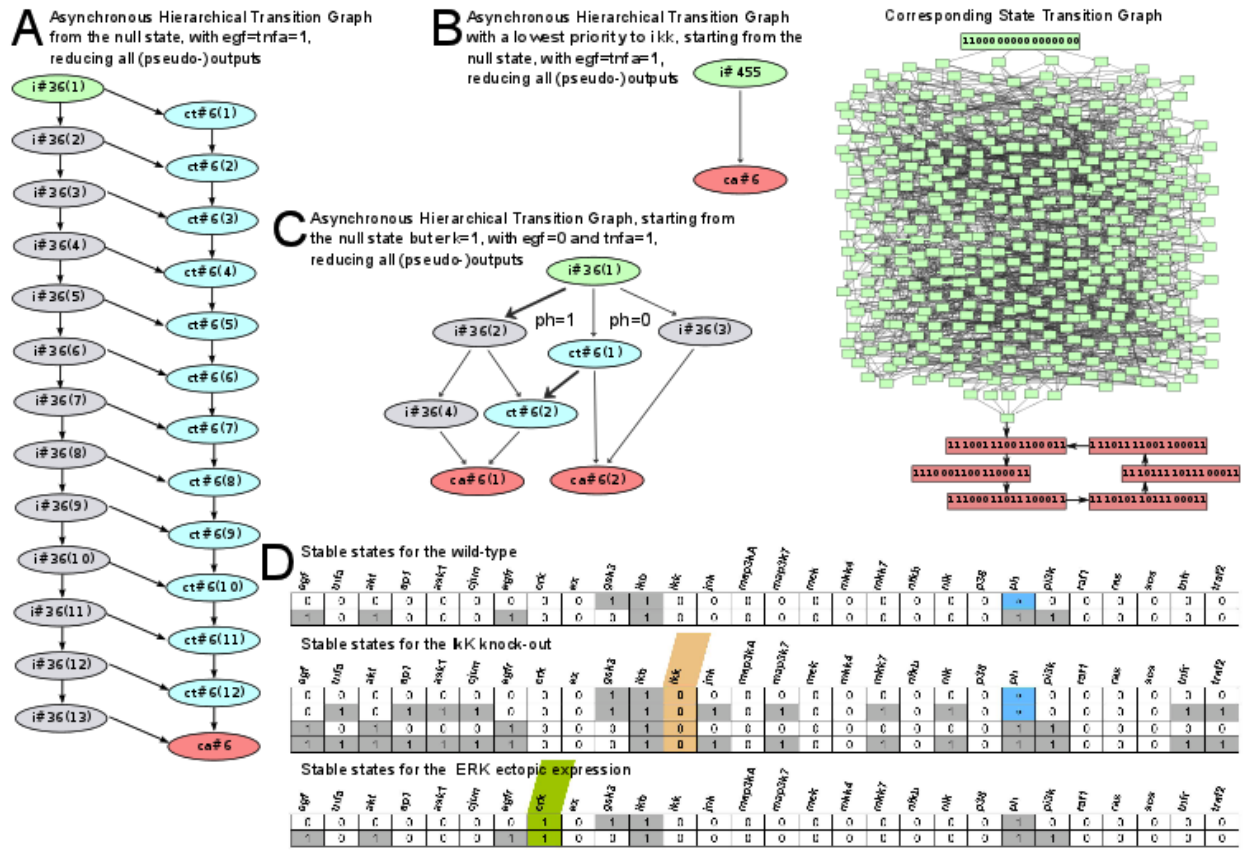

**Figure 5: Properties of the EGF/TNFα model analysed with GINsim.** (A) The Hierarchical Transition Graph (HTG) representing the dynamics of the reduced model (i.e., [pseudo-] outputs removed), under the asynchronous scheme, starting from the null initial state and the EGF=1 & TNFα=1 condition. The HTG shows the organisation of the dynamics with a chain of 13 irreversible sets (each with 36 states) connected, at each stage, to a chain of 12 transient cycles (encompassing 6 states) and a unique cyclic attractor. (B) By defining a lower priority for the update of IKK, all transition states towards transient cycles (in light blue in panel A) are prevented: the system reaches the cyclic attractor without visiting the same state twice. On the right of the panel B, the corresponding State Transition Graph (STG) starting from the initial state (contained in the HTG state set in green) and leading to the cyclic attractor in pink. This STG is shown to illustrate the complexity of the transient dynamics and is not meant to be readable. (C) The HTG showing that, under the asynchronous scheme, different attractors are reachable. Here the two cyclic attractors differ by the value of ph (arrows in bold embody transitions increasing the value of ph). (D) Stable states for the wild-type, IKK knock-out and ectopic expression of ERK, respectively.

## Simulations and biological application with the Cell Collective

The Cell Collective platform aims at facilitating collaborative modelling for experimental scientists. Examples of simulations of the EGF/TNFα model with the Cell Collective are illustrated in Figure 6. Specifically, the Cell Collective offers two modes of simulations: input-output dynamical analyses across hundreds of simulated environments, along with real-time, interactive simulations.

Figure 6A-C illustrates simulations of input-output relationships between external species stimulating the network and species of interest regulated in response to this stimulation. Laboratory studies to identify functional relationships between extracellular stimuli and various cellular components are often expansive and resource consuming. The dynamical analysis tool implemented in the Cell Collective allows users to conduct *in silico* experiments mimicking laboratory experiments, with the advantage that researchers can simulate hundreds or thousands of extracellular and/or disease-related situations (as opposed to the limited number of scenarios that can be reasonably handled in the laboratory) and generate rich input-output relationships (i.e., dose-response curves) between network stimuli and any species in the network. In this respect, inputs and outputs are continuous values on a scale from 0 to 100 (see Methods), despite the discrete (Boolean) nature of the network model. For example, Figure 6A shows a dose-response curve and a positive correlation between EGF and Akt. In contrast, inhibition of PI3K results in the loss of EGF-dependent activation of Akt (Figure 6B). Finally, the input-output relationship between TNFα and IκB is illustrated in Figure 6C.

Real-time interactive simulations with the Cell Collective enable users to interactively change the environment during the simulations. This tool enables users to test "what-if" scenarios, e.g., changes in the external conditions, as well as of (transient) gain/loss-of-function, with instant feedback in terms of the changing activity levels of affected species. To illustrate the utility of this mode, we simulated the EGF/TNFα model under a condition where EGF was set to a medium activity level, while keeping TNFα inactive. This condition results in the activation of Akt, Erk, and Ras (Figure 6D). The simulation of a Ras gain-of-function results in further activation of Erk, but not of Akt (Figure 6E). In contrast, Akt continues to respond to EGF activation and deactivation (Figure 6F and G, respectively). This is because Akt (unlike Erk) does not lie downstream of Ras (Figure 2) and hence is not affected by the constitutively activated Ras.

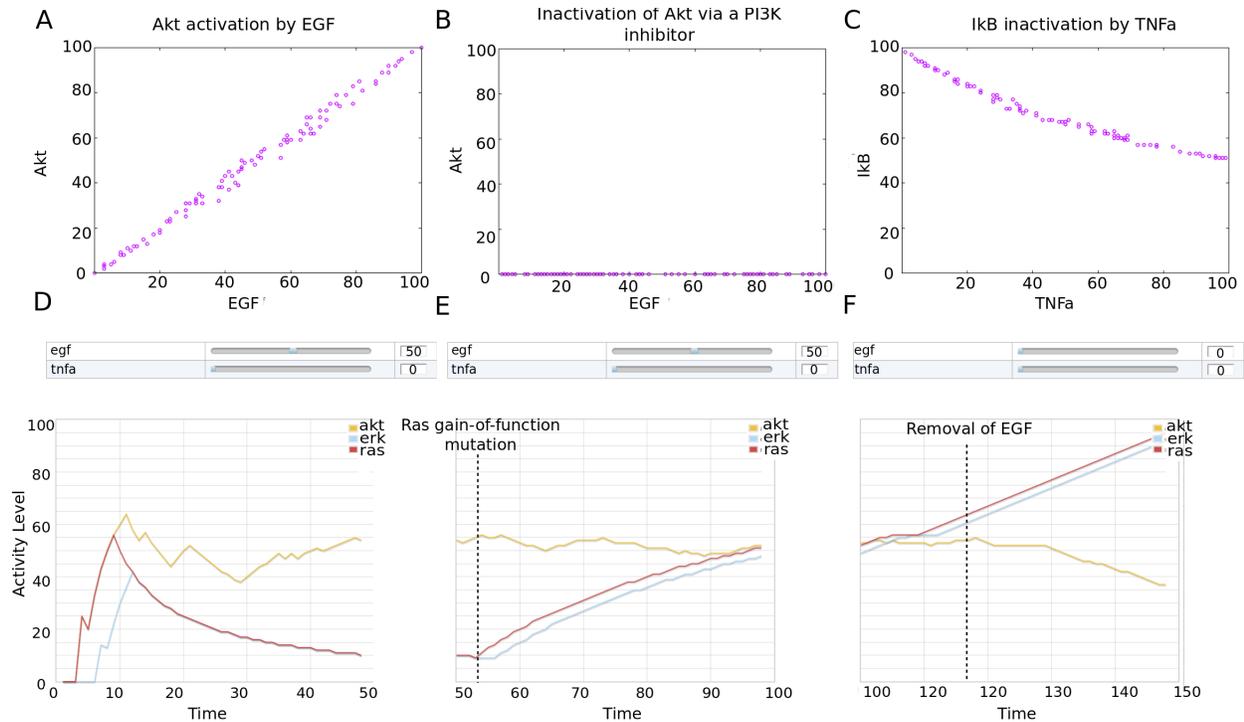

**Figure 6: Examples of simulations in the Cell Collective.** In panels A-C, the model was simulated 100 times, 800 time steps each. For each of the 100 simulations, an activity level between 0 and 100 (i.e., probability of being active at time *t*) was randomly selected for EGF (panels A and B) and TNFα (panel C); these values comprise the abscissa. A) Dose-response curve illustrating the activation of Akt under changing levels of EGF. B) Inhibition of PI3K results in the loss of EGF-induced activation of Akt. C) Dose-response curve illustrating inactivation of IkB in response to increasing levels of TNFα. D-F) Real-time simulation under varying conditions: D) Setting EGF to 50% (using the sliders illustrated above the plot) results in an intermediate activation of Akt, and transient activation of Ras and Erk. E) Simulated Ras gain-of-function (introduced around time step 50), results in the activation of Ras, which subsequently stimulates Erk. Akt remains active at around 50%. F) The removal of EGF (turning it to 0%) results in the decrease of Akt activity, while Erk continues to rise due to Ras mutation. Any and all species in the model can be displayed during the real-time simulations; the three species Akt, Erk, and Ras were selected for illustration purposes only. Note that the activation levels do not correspond to concentrations or any molecular measurements; they rather provide a semi-quantitative activity measure to analyse the effects of changes in the model (e.g., perturbations) on the rest of the network.

## The LogicalModel library

In order to ease the adoption of the new standard, an open source (Java) library, LogicalModel has been created. The library can be used as a standalone command line tool for model conversion, and can be accessed at https://github.com/colomoto/logicalmodel. It provides a data structure to manipulate logical models, as well as a set of analytic tools (e.g., stable state identification, model reduction) that are common to many scientific efforts relying upon a discrete modelling approach. The library further provides import and export filters for SBML qual (through JSBML; see Methods section), as well as an interface enabling the integration of logical models and SBML qual with additional formats and a number of existing software tools. The development of this library coincides with the onset of CoLoMoTo initiative.

Figure 7 illustrates the central role of the LogicalModel library and main current model exchange capabilities of popular software tools, covering logical models as well as additional related qualitative modelling frameworks. Included are importers that have been recently developed to generate (non parameterized) qualitative models from pathway databases: KEGGtranslator [73] and Path2Models project [74].

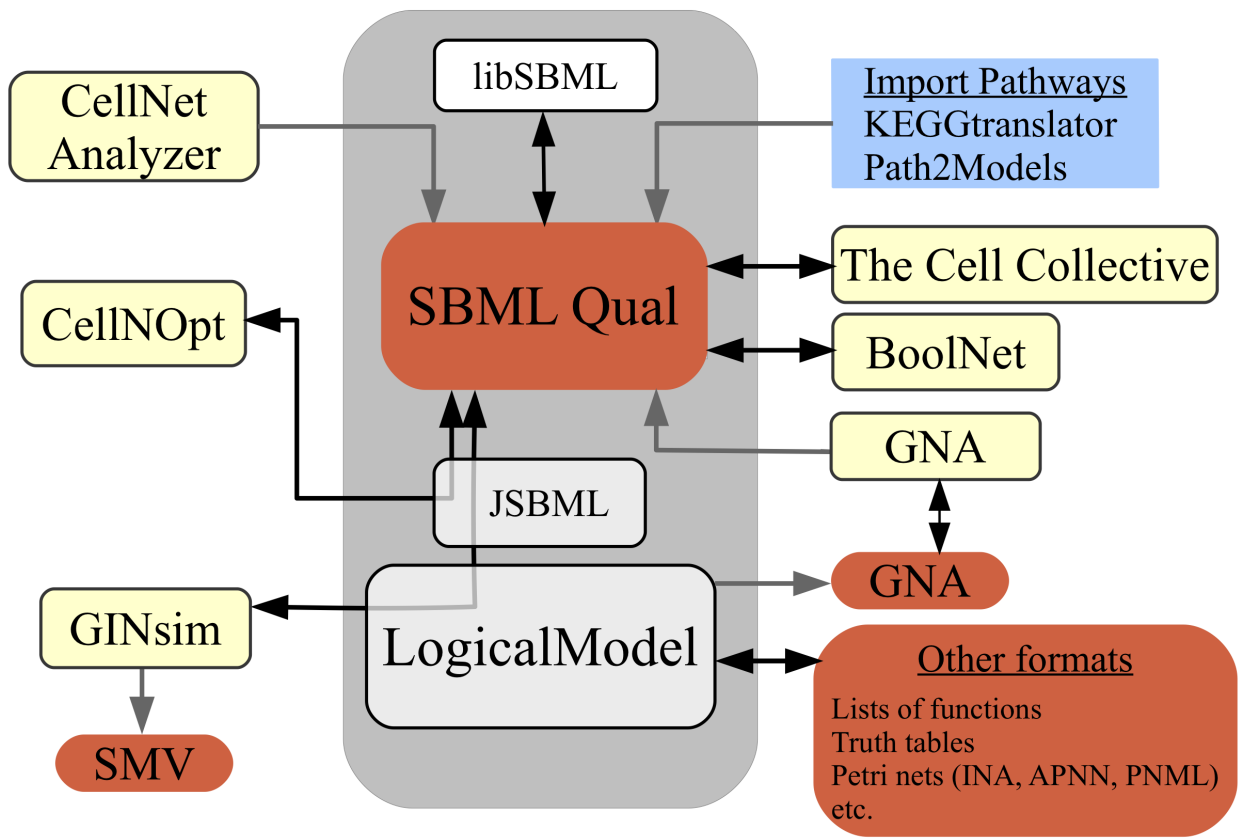

**Figure 7: LogicalModel library as an interface among various qualitative modelling technologies.** Orange boxes denote formats, light yellow boxes are software tools and white boxes include libraries. Arrows denote export/import capabilities.

# Conclusions and prospects

In order to enable the interoperability of qualitative, discrete models, a standard exchange medium was necessary. While the previous versions of SBML were not fully compatible with qualitative modelling approaches, the modular structure of SBML release (Level 3) enables the development of additional packages to support novel modelling frameworks and capabilities.

In this paper, we report on the SBML Level 3 Qualitative Model (SBML Qual) package, which provides a standard means for the exchange of logical models of regulatory and signalling networks. Currently, at least three software tools implement SBML qual (GINsim, CellNOpt, and the Cell Collective), while other tools such as GNA [75] and CellNetAnalyzer can export models to this format. The former three tools have been used here to demonstrate the consistency of the standard via simulations and analyses of a Boolean model of EGF/TNFα signal transduction pathways. The combined use of software tools is now facilitated, providing modellers with a range of complementary means to investigate their models.

Repositories of models encoded in SBML qual are already being prepared. For instance, the Cell Collective now contains numerous previously published logical models that can be downloaded. BioModels, which serves as a reliable repository of computational models of biological networks [76], currently hosts several curated SBML qual models (http://www.ebi.ac.uk/biomodels/).

SBML qual will continue to be refined by the community. Some of the improvements discussed so far include, for example, the definition of models where parameters are not (all) instantiated, models for which timing constraints are specified, extended Petri nets, etc. In addition, further integration with SBML Core concepts is planned. In particular, SBML qual will be useful to so-called hybrid formalisms, which combine features of both discrete and continuous models. A typical example are formalisms embedding a logical representation of the interaction structure of the network into a continuous model of its dynamics, such as piecewise-linear differential equation models [77], hybrid automata [78, 79], or even fully continuous models in which the logical functions have been replaced by sigmoidal functions preserving the logic of the interactions [80]. Other hybrid formalisms that have been used for the modelling of regulatory and signalling networks are fuzzy logic-based models [81] and timed automata [82, 83]. Software tools enabling the modelling, simulation, and analysis of networks by means of different kinds of hybrid models include Odefy [39], SQUAD [41], GNA [75], and Q2LM [84]. Most of the above-mentioned tools support SBML Core.

Last but not least, to further support data and result reproducibility, the standardisation of algorithms and simulation schemes and parameters for qualitative models is planned by adopting the MIASE guidelines [85]. A first step in this direction has, in fact, already been taken by adding simulation methods relevant to logical models to the Simulation Experiment Description Mark-up Language (SED-ML, [86]).

In conclusion, the availability of SBML qual and the inception of the CoLoMoTo consortium should foster the collaborative development of standards (including the extension of existing ones), as well as of computational methods for the qualitative modelling of biological networks. In this respect, anyone interested in these efforts is cordially invited to enter into contact with the existing community at sbml-qual@lists.sourceforge.net.

**Authors' Contributions**

CC, JSR, and TH designed the workflow and case studies presented here and wrote the manuscript. CC initiated, coordinated and contributed to the development of SBML qual, CC coordinated the support of the package in GINsim. JSR contributed to the development of SBML qual, and supervised its implementation in CellNOpt. TH contributed to the development of SBML qual, and coordinated the implementation of its support in the Cell Collective. DT initiated and regularly contributed to the development of SBML Qual; DT also contributed to the redaction of the manuscript. S. Keating contributed towards the finalization of the SBML qual specification and developed the libSBML code that supports the format. DB contributed to the development of the qual package and developed a first support in LibSBML. AN contributed to the qual package, implemented its support in GINsim and implemented the LogicalModel library. MPI implemented an XML validator for SBML qual as part of LibSBML, helped review, collect feedback and improve the SBML qual specification, and created the first implementation of the CellNOpt-to-SBML-qual converter. AD and FB contributed to the development of SBML qual and to JSBML. JD implemented the SBML qual export in boolSim. IX contributed to the development of SBML Qual. EG contributed to the support of SBML qual in CellNOpt and to the manuscript. PTM implemented the GINsim export to GNA, contributed to the GINsim support for SBML qual and to the manuscript. MP and HdJ tested the specification by developing SBML qual support for GNA; HdJ contributed to the manuscript. MH helped guide development of the SBML Level 3 "qual" package specification and contributed to the manuscript. S. Klamt contributed to SBML qual development and supervised its support in CellNetAnalyzer. AVK contributed to the requirements of the SBML package, and implementation of SBML support in CellNetAnalyzer. NR contributed to the development of SBML qual, and implemented support for the specification in JSBML. BK and BW implemented SBML qual support in the Cell Collective. TC participated in the case-study setting-up, simulations and import/export in CellNOpt, and contributed to the manuscript.

**Acknowledgements**

Authors thank the members of the SBML qual mailing list for their contributions and/or feedback during the development of the qual package specification. We also thank Aidan MacNamara for help with the EGF/TNFα model. Furthermore, we appreciate financial aids from the European Union through the "BioPreDyn" project to JSR (ECFP7-KBBE-2011-5 Grant #289434), Nebraska NASA Space Grant Consortium grant ("Technology for collaborative biomedical research") and National Institutes of Health grant support (#5R01DA030962) to TH, Fundação para a Ciência e a Tecnologia (grant PTDC/EIA-CCO/099229/2008) to CC, US National Institute of General Medical Sciences support (#GM070923) for SK, MH and NLN, a Marie Curie International Outgoing Fellowship within the EU 7th Framework Program for Research and Technological Development (project AMBiCon, 332020) to AD, the Swiss Federal Government through the Federal Office of Education Science and Innovation (SERI) and the European Commission FP6 project ENFIN (Experimental Network for Functional INtegration – LSHG-CT-2005-518254) to IX, the Fundação para a Ciência e a Tecnologia (grant Pest-OE/EEI/LA0021) to PTM, Federal Ministry of Education and Research (BMBF, Germany) as part of the Virtual Liver Network (grant number 0315756) to AD, SK. FB, MvI, JSR and NLN also benefited from dedicated support by EMBL-EBI.

# Supplemental Information

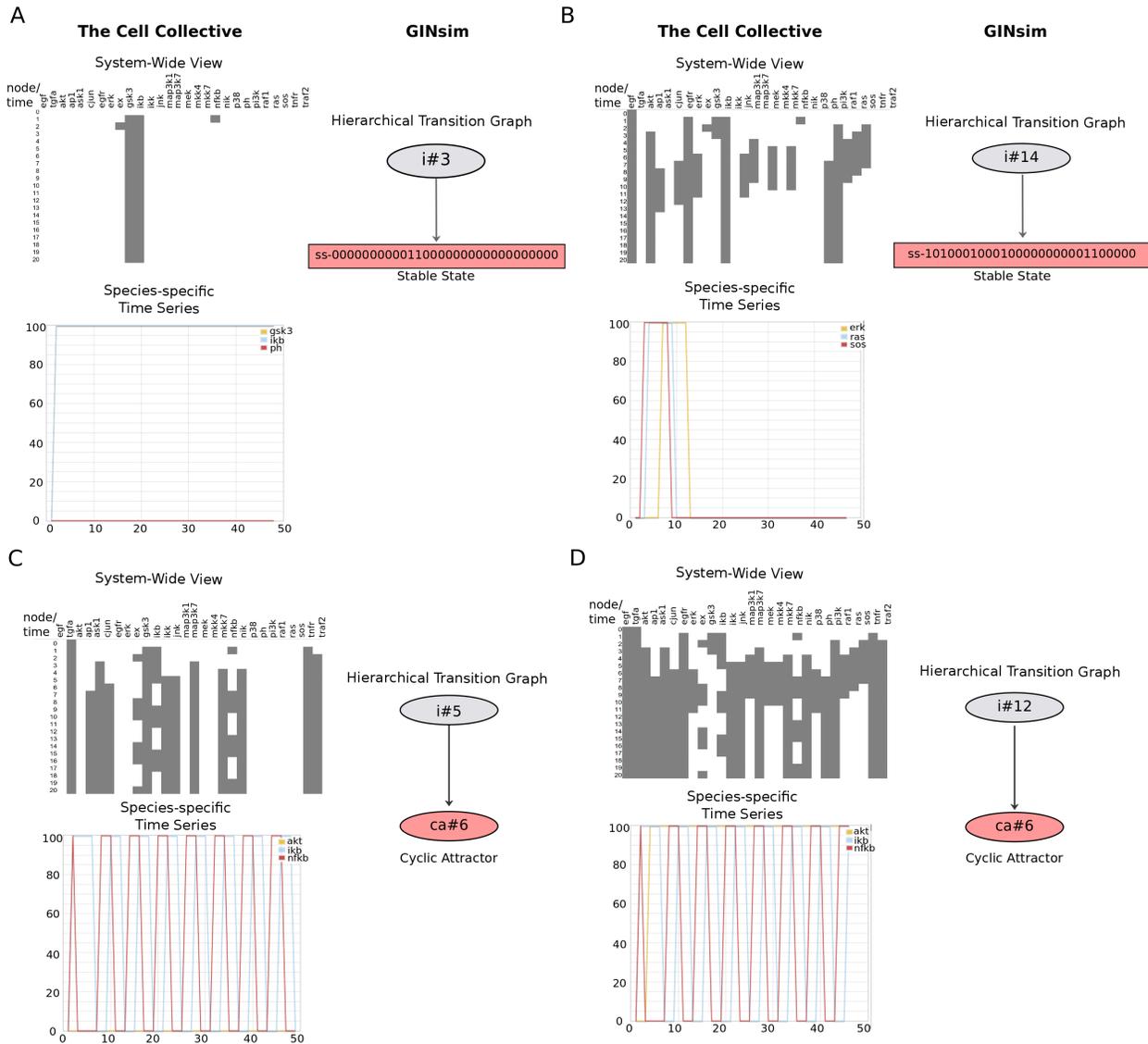

**Figure S1. Complete dynamic profile of the example model.** The model was simulated with consistent results in the Cell Collective (left column of panels A & B), GINsim (right column of panels A & B), and CellNOpt (data not shown, but simulations were consistent with those presented here). Charts at the top of the Cell Collective column correspond to the overall dynamic profile across all nodes in the model. Black cells correspond to active (1) states, whereas inactive (0) states are white. The bottom graphs in the Cell Collective column illustrate the time course of selected nodes. The GINsim columns show State Transition Graph and the Hierarchical Transition Graph (HTG) generated with the tool. Note that due to the synchronous updating, the irreversible components of the (HTG) correspond to linear chains of states. Simulations were performed under four input conditions: EGF=TNFα=0; EGF=TNFα=1; EGF=0 & TNFα=1; EGF=1 & TNFα=0. A) EGF=TNFa=0. The network reaches a steady state (shown in both GINsim and the Cell Collective column) after 3 transient states. The order of the individual species states in the steady state generated by GINsim is sorted in the same (alphabetical) order, as presented in the Cell Collective

column. B) EGF=1, TNFa=0. The network reaches a steady state after 14 transient states. C) EGF=0, TNFa=1. Following 5 transient states, the network reaches a 6-cycle attractor. D) EGF=TNFa=1. After 12 transient states, the network reaches a cyclical attractor encompassing six states.